# A simple model for describing a lattice with a double occupancy

Valentin Voroshilov

Physics Department, Boston University, Boston, MA, 02215, USA

A simple model is presented to investigate an impact of the double occupied sites on the ground state energy of a lattice. The model is seen as a useful tool to introduce undergraduate or graduate physics students to an array of a relatively simple mathematical apparatus and physical ideas which they can choose later for a deeper study. Instead of analyzing a system of electrons in a periodic potential, a system of sites having different energy states related to a number of extra electrons at a site is considered. The simplification is achieved by introducing operators for one-electron sites and two-electron sites as independent entities. A simple modeling function for the ground state of the system is constructed. Linear and quadratic lattices are considered. For a quadratic lattice the existence of a critical temperature is shown.

## Constructing the Hamiltonian

Since the first high temperature superconductor was discovered[1], one of the popular approaches for describing the phenomenon of HTSC involves a Habbard–like Hamiltonian for a system of electrons in a lattice[2,3]. The challenge is analyzing the Hamiltonian having a bi-quadratic term which describes interactions between electrons. Difficulties students experience when encounter the mathematical apparatus might divert them from pursuing the career of a physicist.

In this paper instead of analyzing a Hamiltonian to prove the existence of double occupied sites, the fact of their existence is used as the basis for the future analysis of the properties of such a system. All the calculations are relatively simple and do not exceed the standard calculus course.

In this section, a plausible reasoning is used to introduce a Hamiltonian of a system, the ground state of which is investigated in the next section. In the third section a possible connection to superconductivity is discussed as one of the motives for the future investigation of the model.

Valentin Voroshilov, valbu@bu.edu, Physics Department, Boston University, 590 Commonwealth Ave., Boston, MA 02215

The fourth section presents generalization into a square lattice, where also the existence of a critical temperature is shown.

______________________________

Let us consider a lattice, the atoms of which have a field strong enough to make a bound state with one additional electron (first electron affinity is negative). In this case there are two strong interactions in the system, namely Coulomb repelling between electrons and Coulomb attraction between electrons and sites.

Even if the second electron affinity is positive, the assumption is that the average energy of the Coulomb repelling between electrons is high and there is a tendency for two electrons to occupy the same site.

In addition, the assumption is used that at a low temperature free electrons do not exist, all electrons spend most of the time sitting at the sites, and sites exchange electrons when those tunneling from one site onto another.

Effectively, instead of considering a system of electrons moving in a periodic potential, a lattice is considered which can have empty sites (a neutral atom), as well as sites with one extra electron (a stable negative ion), or sites with two extra electrons at the site (a negative ion). Under the described conditions, when the number of electrons $N_e$ is large enough relative to the number of sites $N$ and the temperature is low enough, some of the sites must have two electrons on them.

Let $V_e$ be the average energy per an electron of Coulomb repelling of electrons which do not belong to the same site; $-E'$ be the energy of a site having one additional electron (for any spin component of $\sigma = +, -$); and $-2E' + U$ be the energy of a site having two extra electrons ($U$ reflects the repelling energy between two electrons sitting at the same site).

For the described lattice, at $T = 0$, if the number of atoms $N$ is rather grater then the number of electrons $N_e$, $N \gg N_e$, all electrons should be expected occupying sites, with each site having no more then one electron; hence, for a rough estimation for the energy of the ground state in this case we can write

$$E_0^< = V_e N_e - E' N_e = - E N_e \qquad (E = E' - V_e > 0) \qquad (1)$$

However, for the case $N < N_e$, at $T = 0$ some of the electrons must occupy the same atom (the assumption is that this is energetically more beneficial for the system then having free electrons), hence a double occupancy naturally exists in this case. When the number of double occupied sites does not change significantly, the Coulomb repelling energy per an electron can be taken as a constant of the system. The rough estimation for the energy of the ground state of the system, when $N < N_e$ and $N_e - N$ sites have two extra electrons and the rest of the sites have one extra electron, can be written as

$$E_0^> = V_e N_e + (-2E' + U)(N_e - N) - E'(N - (N_e - N)) = -UN + (-E + U) N_e \qquad (2)$$

Our goal is to write a Hamiltonian for the system of the sites in terms of operators describing the neutral sites, as well as the sites having a single extra electron and a pair of extra electrons.

The sites can change their states by exchanging electrons. Further, only electron exchanges (transitions) between the nearest sites are considered, and the transitions which involve only one electron and happen without a change of the spin component of the electron. In addition, an assumption is made that direct transitions and reversed ones have the same probability, which means that direct and reversed transitions make the same input into the exchange energy of the system.

In the first three sections a linear lattice with a periodic boundary condition is considered.

A traditional approach for investigating properties of a many-body system is based on analyzing a Hamiltonian written in terms of creation and annihilation operators. Because we want to consider explicitly the existence of sites having one electron (analog of single electrons in the lattice), as well as sites having two electrons (analog of pairs of electrons), two types of operators are to be introduced.

Each site might have four possible states; hence, the Hilbert space of a site has four vectors. Let $|i; 0, 0>$ be the vacuum state (no single extra electron, no pair of extra electrons) of an $i$-th site (i.e. an $i$-th site is a neutral atom); let $|i; +, 0>$ be the state with an extra electron with the spin up, and $|i; -, 0>$ be the state with an extra electron with the spin down; let $|i; 0, 1>$ be the state with a pair. The state vector of the system is a tensor product of the state vectors of the sites.

Let $a_{i\sigma}^+$ and $a_{i\sigma}$ be operators acting on the one-electron part of the states of a site $i$; and $b_i^+$ and $b_i$ be operators acting on the two-electron part of the states of a site $i$; $i$ numerates sites (atoms, ions) of the lattice; $\sigma = +, -$. Notice, that the $b$-operators are not constructed from the $a$-operators[4]; for example, $b^+ \neq a_+^+ a_-^+$; within this simple model the $b$-operators are independent entities.

Let us define $a_\sigma^+$ and $a_\sigma$; $b^+$ and $b$ operators by the following actions (here and further the indication of a site is omitted when it does not lead to misunderstanding).

$$a_\sigma^+|0,0> = |\sigma,0> \quad a_+|+,0> = |0,0> \quad a_-|-,0> = |0,0> \quad a_\sigma|0,0> = 0 \quad (3a)$$

$$a_+|-,0> = 0 \quad a_-|+,0> = 0 \quad a_+^+|+,0> = 0 \quad a_-^+|-,0> = 0 \quad (3b)$$

$$a_-^+|+,0> = 0 \quad\quad\quad a_+^+|-,0> = 0 \quad (4)$$

$$b^+|0,0> = |0,1> \quad b|0,1> = |0,0> \quad b^+|0,1> = 0 \quad b|0,0> = 0 \quad (5)$$

$$a_\sigma^+|0,1> = 0 \quad a_\sigma|0,1> = 0 \quad b^+|\sigma,0> = 0 \quad b|\sigma,0> = 0 \quad (6)$$

The relationships (3) and (5) are similar to a standard definition of the Fermi creation and annihilation operators. The relationships (4) and (6) show that the *a*-operators do not have any influence on states with a pair, and, in turn, the *b*-operators do not have any influence on single electron states. In terms of electrons, we could say that there is no state such as two electrons sitting at a site, instead of that there is a state of a pair of electrons sitting at a site. All the provided in (3) – (6) relationships based on the assumption that a site can have only four possible states, namely, a site does not have any extra electrons on it; a site have one extra electron with a spin up, a site have one extra electron with a spin down; a site have one pair of extra electrons (a singlet); but a state $|\sigma, 1>$ does not exist, as well as a state $|+ \text{ and } -, 0>$ does not exist ether, and free electrons also do not exist in the system.

The commutative and anti-commutative relationships for the *a*- and *b*-operators assigned to the same site can be derived directly from the definitions (3) – (6) and provided (partially) in the Appendix (and do not agree with canonical Fermi or Bose relationships).

Because operators assigned to different sites act on the states of the sites independently, we prescribe them commutative relationships when they do not belong to the same site ($i \neq j$).

Now the Hamiltonian of the system can be introduced.

In this simple model, the potential energy of the system relative to the average repelling energy of electrons is produced by the sites having one or two electrons. Hence, let us write the Hamiltonian of the system as

$$H = H_{int} + H_{exc} \quad (7a)$$

$$H_{int} = -E \sum_{i\sigma} a_{i\sigma}^+ a_{i\sigma} + (-2E + U) \sum_i b_i^+ b_i \quad (7b)$$

Here, $H_{int}$ represents the energy contributed by the sites and has the structure of (2); $H_{exc}$ represents the exchange energy of the system, and the summation runs over all the sites of the lattice and spin components of extra electrons.

To write the exchange energy term we need to reflect on the transitions electrons (and pairs, in a general case) can make between the sites.

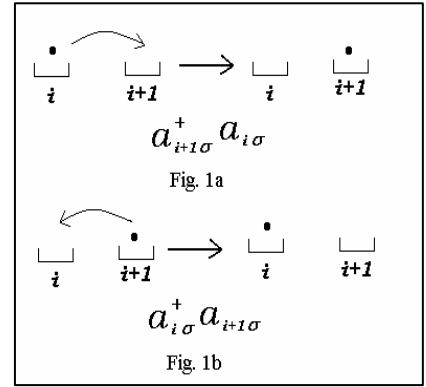

Fig. 1a

Fig. 1b

In the Figure 1a the easiest one-electron transition is shown, when an electron is tunneling from the *i*-th site onto the *i*+1-th site, which previously was empty (neutral); also the correspondent term in the Hamiltonian is written. In the Figure 1b the reversed process is shown.

In the Figure 2a another possible one-electron transition is shown, when an electron is transferring from the *i*-th site onto the *i*+1-th site, which previously was not empty; hence, after this transition a pair of electrons is created on the *i*+1-th site. In the Figure 2b the reversed process is shown, when a pair breaks up into two single electrons.

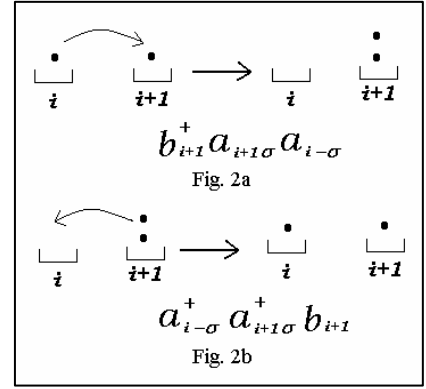

Fig. 2a

Fig. 2b

However, the process shown in the Figure 2a has another possible reversed transition, shown in the Figure 2c, when an electron moves from *i*+1-th site onto *i*-th site, and makes a pair on that site. This process has, in turn, its own reversed transition, shown in the Figure 2d. Therefore, there are four more terms with the same probability which are supposed to be included into the exchange energy of the system.

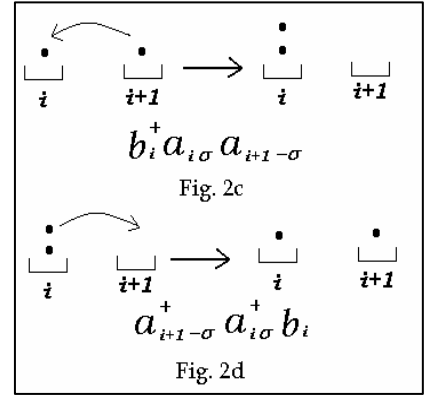

Fig. 2c

Fig. 2d

Finally, there is one more possible process involving one electron, which is shown (and its reversed process) in the Figures 3a and 3b.

Now the exchange energy term of the Hamiltonian can be written.

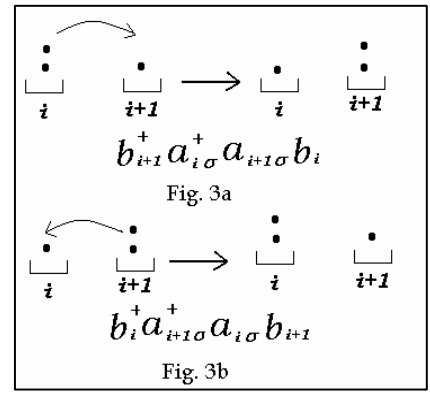

Fig. 3a

Fig. 3b

$$H_{exc} = \eta \sum_{i\sigma}(a^+_{i+1\sigma}a_{i\sigma} + a^+_{i\sigma}a_{i+1\sigma}) + \quad (8)$$

$$+ \tau \sum_{i\sigma}(b^+_{i+1}a_{i+1\sigma}a_{i\,-\sigma} + a^+_{i\,-\sigma}a^+_{i+1\sigma}b_{i+1} + b^+_i a_{i\sigma}a_{i+1\,-\sigma} + a^+_{i+1\,-\sigma}a^+_{i\sigma}b_i)$$

$$+ t' \sum_{i\sigma}(b^+_{i+1}a^+_{i\sigma}a_{i+1\sigma}b_i + b^+_i a^+_{i+1\sigma}a_{i\sigma}b_{i+1})$$

At a very low temperature and when $N_e > N$, transitions shown in the Figures 1 and 2 should be suppressed relative to the transitions shown in the Figures 3. The latter transitions happen without a change in the structure of the state of the system (the number of sites having one extra electron or an extra pair stays the same); the transitions shown in the Figures 2 change the structure of the state; the transitions shown in the Figures 1 are impossible to exist at the zero temperature, because in the case of $N_e > N$ the ground state is not supposed to have any empty sites. Hence, we can expect that at a low temperature the last term in the exchange energy dominates over the others. Assuming that the exchange processes lower the energy of the system we introduce a positive parameter $t = -t'$.

Keeping in (8) the last term only and combining (7) and (8) (and exchanging $a$-operators in the last exchange energy term) we obtain the Hamiltonian of the system in the following form.

$$H = -t \sum_{i\sigma}(b^+_{i+1}a_{i+1\sigma}a^+_{i\sigma}b_i + b^+_i a_{i\sigma}a^+_{i+1\sigma}b_{i+1}) - E \sum_{i\sigma} a^+_{i\sigma}a_{i\sigma} + (-2E + U)\sum_i b^+_i b_i \quad (9)$$

Even though, when $N_e < N$ the system might have empty sites, we still assume that at a low temperature the transitions shown in the Figures 3 are the most probable, hence the Hamiltonian (9) will be used in both cases.

The next step is investigating the ground state of the Hamiltonian (9).

For investigating the ground state of the system with a fixed number of extra electrons the operator (10) is to be introduced.

$$Ne = \sum_{i\sigma} a_{i\sigma}^+ a_{i\sigma} + 2\sum_i b^+_i b_i \quad (10)$$

This operator counts in a natural way the sites having one or two extra electrons.

We should emphasize that the Hamiltonian (9) is obtained purely from plausible reasoning, and the long sequence of assumptions ahs been made, hence at this stage one cannot use the Hamiltonian (9) to make solid physical conclusions.

## Estimating the Ground State Energy

Because *a*- and *b*-operators do not obey canonical commutative relationships, the regular apparatus developed for Bose operators cannot be used. However, in the matrix representation, the matrixes for the operators are rather simple (see the Appendix) and direct calculations can be done.

The vacuum state of the system can be written as the tensor product of the vacuum states of sites:

$$|vac> = |1; 0, 0>|2; 0, 0>|3; 0, 0>|4; 0, 0>...|N-1; 0, 0>|N; 0, 0> = \prod_{i=1}^{N} |i; 0, 0> \quad (12)$$

To estimate the value of the energy of the ground state a modeling state vector for the ground state of the system is constructed.

First, let us consider the case of $N_e < N$. The ground state energy expectation value is anticipated to be given by (1).

It is naturally to expect for the system with the Hamiltonian (9) that, when the number of extra electrons is less than the number of the sites, each site is occupied with one extra electron only. Because the existence of pairs in the ground state with $N_e < N$ is not expected, it is naturally to introduce the following state vector for the ground state of the system, where *u* and *v* are real numbers

$$|E_0^<> = \{u(a_{N+}^+ + a_{N-}^+) + v1_N\}...\{u(a_{1+}^+ + a_{1-}^+) + v1_1\}|vac> =$$

$$= \prod_{i=1}^{N} \{u(a_{i+}^+ + a_{i-}^+) + v1_i\} \prod_{i=1}^{N} |i; 0, 0> \quad (13)$$

The parameter *u* represents the probability amplitude for a site of having a single extra electron with the spin up or down, and the parameter *v* represents the probability amplitude for a site of being empty.

The condition $<E_0^<|E_0^<> = 1$ leads to

$$2u^2 + v^2 = 1 \tag{14}$$

The ground state energy can be estimated as

$$E_0^< = <E_0^<|H|E_0^<> \tag{15}$$

In (15) $E_0^<$ is a function of the parameters $u$ and $v$. To find the parameters, the condition that the total number of electrons $N_e$ is a given constant is applied.

$$<E_0^<|Ne|E_0^<> = N_e \tag{16}$$

With the use a the matrix representation, it is easily seen that

$$<E_0^<|a^+_{i\sigma} a_{i\sigma}|E_0^<> = u^2 \tag{17a}$$

$$<E_0^<|b_i^+ b_i|E_0^<> = 0 \qquad <E_0^<|b_{i+1}^+ a_{i+1\sigma} a_{i\sigma}^+ b_i|E_0^<> = 0 \tag{17b}$$

$$E_0^< = -2ENu^2 \qquad N_e = 2u^2 N \tag{17c}$$

The formula (1) is easily obtained from the relations (17c) and can be written in a slightly different form

$$E_0^</N = -Ex \qquad N_e/N = x \qquad (x < 1) \tag{18}$$

Here and further x is an electron density (the average number of electrons per a site).

Now let us consider the case of $x > 1$ ($N_e > N$).

When the system with the Hamiltonian (9) is in the ground state, it is naturally to expect that each site might be represented by an ion having a single additional electron with the spin up or down, or by an ion having a pair of additional electrons. Hence, it is naturally to construct the modeling ground state vector as

$$|E_0^>> = \{u(a_{N+}^+ + a_{N-}^+) + wb_N^+\}...\{u(a_{1+}^+ + a_{1-}^+) + wb_1^+\} |vac> =$$

$$= \prod_{i=1}^{N} \{u(a_{i+}^+ + a_{i-}^+) + wb_i^+\} \prod_{i=1}^{N} |i; 0, 0> \tag{19}$$

The new parameter $w$ represents the probability amplitude for a site of having a pair of electrons.

The condition $<E_0^>|E_0^>> = 1$ leads to $2u^2 + w^2 = 1$.

When calculating $E_0^> = <E_0^>|H|E_0^>>$ and $N_e = <E_0^>|Ne|E_0^>>$, the result is

$$<E_0^>| a^+_{i\sigma} a_{i\sigma}| E_0^>> = u^2 \quad <E_0^>| b_i^+ b_i| E_0^>> = w^2 \quad <E_0^>| b_{i+1}^+ a_{i+1\sigma} a_{i\sigma}^+ b_i |E_0^>> = u^2 w^2 \quad (20a)$$

$$N_e = 2Nu^2 + 2Nw^2 \quad u^2 = 1 - \tfrac{1}{2}x \quad w^2 = x - 1 \quad x = N_e/N \quad (x > 1) \quad (20b)$$

$$E_0^>/N = -U + (-E + U)x + 2t(x^2 - 3x + 2) \quad \text{or} \quad (21a)$$

$$E_0^>/tN = -xE/t + \delta x + 2x^2 - 6x + 4 - \delta \quad \delta = U/t \quad (21b)$$

When $t = 0$ (21a) becomes (2), which is a natural result since the exchange energy of the system was not taken into a consideration when writing (2).

## Possible Connection to Superconductivity

Since the height temperature superconductivity is a very exiting phenomenon, a discussion connecting the model to the properties of height temperature superconductors might be a good motivational tool to support students' interest to contemporary physics.

The existence of double occupied sites does not necessarily mean the existence of superconductivity.

As a possible bridge between the double occupancy and superconductivity within this model we consider the existence of *additional* electron pairs. For example, if for the case of $x < 1$ we could find a state having an energy lower than (18) but at the same time having pairs of extra electrons; or, if for the case of $x > 1$ we could find a state having energy lower than (21a) but at the same time having more pairs of extra electrons than the state (19), we could consider the existence of those extra pairs as the sing for superconductivity, because the formation of additional electron pairs might be considered as a spontaneous quantum effect.

The natural way to introduce a state vector for the state which could exhibit the existence of extra electron pairs is

$$|E_0> = \{u(a_{N+}^+ + a_{N-}^+) + v1_N + wb_N^+\}...\{u(a_{1+}^+ + a_{1-}^+) + v1_1 + wb_1^+\} |vac> =$$

$$= \prod_{i=1}^{N} \{u(a_{i+}^+ + a_{i-}^+) + v1_i + wb_i^+\} \prod_{i=1}^{N} |i; 0, 0> \quad (22)$$

The calculation completely similar to the previous one leads to the following formulae

$$<E_0| a^+_{i\sigma} a_{i\sigma}| E_0> = u^2 \quad <E_0| b_i^+ b_i| E_0> = w^2 \quad <E_0| b^+_{i+1} a_{i+1\sigma} a^+_{i\sigma} b_i |E_0> = u^2 w^2 \quad (23a)$$

$$2u^2 + v^2 + w^2 = 1 \qquad N_e = 2Nu^2 + 2Nw^2 \qquad (23b)$$

$$u^2 = \tfrac{1}{2}x - w^2 \qquad v^2 = 1 - x + w^2 \qquad x = N_e/N \qquad (23c)$$

$$E_0/N = <E_0|H|E_0>/N = -Ex + Uw^2 - 2txw^2 + 4tw^4 \qquad (23d)$$

By minimizing (23d), the parameter $w^2$ can be found as

$$w^2 = x/4 - \delta/8 \qquad (\delta = U/t) \qquad (24)$$

That leads to the ground state energy per a site

$$E_0/N = -Ex - t(x - \delta/2)^2/4 \qquad (25)$$

Because the $u$, $v$, $w$ parameters must be real numbers, and for having the extremum of the function $E_0(w^2)$ being a minimum, there are additional conditions on the set of the variables involved into the analysis

$$t > 0 \qquad \delta/2 < x < 4/3 - \delta/6 \qquad (26)$$

Now, let us consider the cases of $x < 1$ and $x > 1$ to compare the energies $E_0$, $E_0^<$, and $E_0^>$.

Let us set $x < 1$. In this case, from (26) we have also the condition $\delta < 2$.

Let us calculate the energy difference $\Delta E^</Nt = (E_0 - E_0^<)/Nt$.

From (1) and (25) we obtain $\Delta E^</Nt = -(x - \delta/2)^2/4$ which is a negative number for all $x$. Hence, if $\delta < 2$ and $\delta/2 < x < 1$ the state (22) is energetically more beneficial relative to the state (13), and the ground state (13) is unstable to a creation of additional electron pairs. To characterize such instability, when $x < 1$, the parameter $w^2 \ne 0$ can be used as the order parameter.

Let us set $x > 1$. The condition $\delta < 2$ still is to be held, and the region for the possible values of $x$ now is $1 < x < 4/3 - \delta/6$.

For the energy difference $\Delta E^>/Nt = (E_0 - E_0^>)/Nt$, from (21c) and (25) we obtain the result of $\Delta E^>/Nt = -(3x/2 - 2 + \delta/4)^2$, which is again a negative number for any $x$; hence the state (22) is again energetically more beneficial relative to the state (19). We can say that the ground state (19) is unstable to a creation of *additional* electron pairs. When $x > 1$, the difference between the parameters $w^2 = x/4 - \delta/8$ (the state (25)) and $w^2 = x - 1$ (the state (19)) can be used as the order parameter, and it is seen that the order parameter in this case is $1 - 3x/4 - \delta/8 = v^2$, which represents the density of the empty sites (neutral atoms).

An interesting fact is that when the parameter $t$ approaches zero the energies $E_0^<$ (18) and $E_0^>$ (21a) show a regular behavior, while the ground state energy $E_0$ (25) has a singularity at $t = 0$. This might be a sign for a perturbation theory is not applicable for the Hamiltonian (9) when $t$ (or $\delta$) is a small parameter of the system.

## Square Lattice

In this section the generalization into a two-dimensional lattice is presented.

Let us numerate the sites of a $NxN$ lattice with $i, j = 1, N$ and write the Hamiltonian

$$H = -t \sum_{i,j,\sigma}(b^+_{i+1j} a_{i+1j\sigma} a^+_{ij\sigma} b_{ij} + b^+_{ij} a_{ij\sigma} a^+_{i+1j\sigma} b_{i+1j} + b^+_{ij+1} a_{ij+1\sigma} a^+_{ij\sigma} b_{ij} + b^+_{ij} a_{ij\sigma} a^+_{ij+1\sigma} b_{ij+1}) - \\ - E \sum_{i,j,\sigma} a^+_{ij\sigma} a_{ij\sigma} + (-2E + U) \sum_{i,j} b^+_{ij} b_{ij} \tag{27}$$

The ground state vector is a tensor product

$$|E_0> = \prod_{i=1}^{N}\prod_{j=1}^{N}\{u(a^+_{ij+} + a^+_{ij-}) + v1_{ij} + wb^+_{ij}\} \prod_{i=1}^{N}\prod_{j=1}^{N}| i; 0, 0 > \tag{28}$$

The operator for the number of electrons is

$$Ne = \sum_{i,j,\sigma} a_{i\sigma}^+ a_{i\sigma} + 2\sum_{i,j} b_i^+ b_i \tag{29}$$

Calculation gives us

$$<E_0|E_0> = 1 \qquad E_0 = <E_0|H|E_0> \qquad N_e = <E_0|Ne|E_0> \tag{30a}$$

$$<E_0| a^+_{ij\sigma} a_{ij\sigma}| E_0> = u^2 \quad <E_0| b_{ij}^+ b_{ij}| E_0> = w^2 \quad <E_0| b^+_{i+1j} a_{i+1j\sigma} a^+_{ij\sigma} b_{ij} |E_0> = u^2 w^2 \tag{30b}$$

$$2u^2 + v^2 + w^2 = 1 \qquad N_e = 2N^2 u^2 + 2N^2 w^2 \tag{30c}$$

$$u^2 = \tfrac{1}{2}x - w^2 \qquad v^2 = 1 - x + w^2 \qquad x = N_e/N^2 \tag{30d}$$

$$E_0/N^2 = -Ex + Uw^2 - 4txw^2 + 8tw^4 \tag{30e}$$

To find the energy $E_0^<$ we set $w = 0$ in the relations (30c-e); that leads to

$$E_0^</N^2 = -Ex \tag{31}$$

To find the energy $E_0^>$ we set $v = 0$ and obtain

$$E_0^>/N^2 = -Ex + Ux - U - 12tx + 8t + 4tx^2 \qquad (32)$$

To calculate the lowest energy, the energy (30e) is minimized over $w^2$, which leads to

$$w^2 = x/4 - \delta/16 \qquad (\delta = U/t) \qquad (33)$$

This gives the ground state energy

$$E_0/tN^2 = -Ex/t - x^2/2 + x\delta/4 - \delta^2/32 \qquad (34)$$

The condition $t > 0$ ensures that the extremum of the function $E_0(w^2)$ is being a minimum. Also, because the $u, v, w$ parameters must be real numbers, the following condition is to be held

$$\delta/4 < x < 4/3 - \delta/12 \qquad (35)$$

Now let us consider the cases $x < 1$ and $x > 1$ to compare the energies $E_0$, $E_0^<$, and $E_0^>$ (in both cases it leads to an additional condition $\delta < 4$).

The energy difference $\Delta E^</Nt = (E_0 - E_0^<)/Nt = -(x - \delta)^2/32$ is negative, as well as the energy difference $\Delta E^>/Nt = (E_0 - E_0^>)/Nt = -(3x - 4 + \delta/4)^2/2$.

It is seen that the state (28) again is energetically the most beneficial; hence, the other states are unstable relative to creating extra double occupied sites.

To have an insight into a behavior of the system at $T > 0$ the free energy of the system can be estimated. Let's assume that when $T > 0$ the average energy of the system still can be written as (30e), where the parameter $w^2$ represents the probability for a site of having a pair of extra electrons, and depends upon the temperature T. The equilibrium value of $w^2$ can be found by minimizing the free energy of the system, to construct which we need to know also the entropy of the system.

Let us consider the state of the system which is the closes to an equilibrium; i.e. all the empty sites, one-electron sites, and two-electron sites are distributed as uniformly as possible, and there are exactly $N_e/2$ electrons with the spin up and down. To find the statistical weight of this state the total number of microscopically unequal states has to be found. Because the locations of $N_1$ one-electron sites and $N_2$ two-electron sites are already set, the statistical weight is equal to the number of combination one can assign to $N_1/2$ out of $N_1$ sites having electron with the spin up (this condition exhausts all the combinations related to electrons with the spin down; and subsystems of empty and two-electron sites do not make an input into the statistical weight of the system). That leads to the entropy of the system ($N_1 \gg 1$; $N_1 + 2N_2 = N_e$; $w^2 = N_2/N^2$)

$$S = \ln\left(\frac{N_1!}{(N_1/2)!(N_1 - N_1/2)!}\right) \approx N_1 \ln 2 = (N_e - 2N_2)\ln 2 = (N_e - 2w^2 N^2)\ln 2 \quad (36)$$

By combining relationships (30e) and (36) and minimizing the free energy per a site of the system $F/N^2 = E/N^2 - TS/N^2$ we obtain the solution

$$w^2 = \frac{x}{4} - \frac{\delta}{16} - \frac{\ln 2}{8t}T \quad (37)$$

At $T = 0$ this expression is consistent with (33), and it leads to an existence of a critical temperature $T^* = 8tw_0^2/\ln 2$ ($w_0^2 = w^2_{T=0}$), at which "the order parameter (for the case of $x < 1$)" $w^2$ becomes zero.

For the case of $x > 1$ the parameter $v^2 = (N^2 - N_1 - N_2)/N^2$ shows the probability for the site of being empty and plays the role of the order parameter, hence the free energy of the system should be rewritten in terms of $v^2$ and then minimized, which leads to

$$v^2 = 1 - \frac{\delta}{16} - \frac{3}{4}x - \frac{\ln 2}{8t}T \quad (38)$$

The critical temperature exists again in this case and found as $T^* = 8tv_0^2/\ln 2$ ($v_0^2 = v^2_{T=0}$).

The fact of the existence of a critical temperature in this simple model is promising and promotes a further investigation of properties of the Hamiltonian (9).

The next immediate and merely technical step for students is analyzing the properties of the system derivable from the ground state energy and the free energy of the system and investigating the influence on them of the other terms included in the exchange energy term (8).

---

At this stage of the investigation one cannot say if the model is related to superconductivity, or magnetic properties of cuprates, or it is just an artificial construct. The Hamiltonian (9) is too simple to deliver the properties of superconducting cuprates, for example the condition (35) does not reflect correctly the connection between the doping and superconductivity (although, it is worthwhile to investigate if the limits on the electronic density per a site x are connected to a possible quantum phase transition in the system). However, further speculations might be an additional motivation to attract students to a deeper study of the related physics.

At the low doping (when $|x - 1| << 1$) the double occupied sites may be seen as defects in the magnetic structure of the cuprates. Superconductive regime might start when the number of the double occupied site reaches some critical value. If one would be trying to describe properties of high temperature superconductors in terms of the Hamiltonian (9), one could think of the oxygen atoms as of the zones of attraction similar to ones created by phonons in the canonical BCS-model of superconductivity, however permanent. The rest of the lattice would be functioning as a vendor of electrons for the electron–oxygen subsystem, and also as a source for an additional background field, and might also participate as a mediating agent when electrons are moving from one oxygen atom to another. The $CuO_2$ plains could be considered as a composition of two lattices; the lattice of oxygen atoms and ions, which is directly responsible for superconductivity, and the lattice of other ions and atoms (if this picture were correct, simple materials like $CuO_2$ or $CuHSO_4$, when doped, also were showing superconductive properties).

Although the Hamiltonian (9) is not derived from any first quantization Hamiltonian, the described model can be seen also as a step to the Habbard model (a Habbard-like Hamiltonian can be written from (9) by replacing the $b$-operators with products of two $a$-operators, and a (28)-like vector can be used to model the ground state vector of the system).

## Appendix. Some properties of the a- and b-operators

It is convenient to use a matrix representation for operators to investigate their commutative and/or anti-commutative properties. To go over to the matrix representation let us introduce in the Table 1 four vectors forming the basis of the Hilbert space of the system at each site.

**Table 1**

$$|0, 0> = |1> = \begin{pmatrix} 1 \\ 0 \\ 0 \\ 0 \end{pmatrix} \quad |+, 0> = |2> = \begin{pmatrix} 0 \\ 1 \\ 0 \\ 0 \end{pmatrix} \quad |-, 0> = |3> = \begin{pmatrix} 0 \\ 0 \\ 1 \\ 0 \end{pmatrix} \quad |0, 1> = |4> = \begin{pmatrix} 0 \\ 0 \\ 0 \\ 1 \end{pmatrix}$$

The matrix elements $O_{ij}$ of an operator $O$ are to be calculating as $O_{ij} = <i|O|j>$.

Combining the definitions of the states (Table 1) with the definitions (3) – (6) we find all the matrix elements for the $a$- and $b$-operators. The Table 2 provides the matrixes for the operators.

**Table 2**

$$a^{+}_{+} = \begin{pmatrix} 0 & 0 & 0 & 0 \\ 1 & 0 & 0 & 0 \\ 0 & 0 & 0 & 0 \\ 0 & 0 & 0 & 0 \end{pmatrix} \quad a^{+}_{-} = \begin{pmatrix} 0 & 0 & 0 & 0 \\ 0 & 0 & 0 & 0 \\ 1 & 0 & 0 & 0 \\ 0 & 0 & 0 & 0 \end{pmatrix} \quad B^{+} = \begin{pmatrix} 0 & 0 & 0 & 0 \\ 0 & 0 & 0 & 0 \\ 0 & 0 & 0 & 0 \\ 1 & 0 & 0 & 0 \end{pmatrix}$$

$$a_{+} = \begin{pmatrix} 0 & 1 & 0 & 0 \\ 0 & 0 & 0 & 0 \\ 0 & 0 & 0 & 0 \\ 0 & 0 & 0 & 0 \end{pmatrix} \quad a_{-} = \begin{pmatrix} 0 & 0 & 1 & 0 \\ 0 & 0 & 0 & 0 \\ 0 & 0 & 0 & 0 \\ 0 & 0 & 0 & 0 \end{pmatrix} \quad B = \begin{pmatrix} 0 & 0 & 0 & 1 \\ 0 & 0 & 0 & 0 \\ 0 & 0 & 0 & 0 \\ 0 & 0 & 0 & 0 \end{pmatrix}$$

Now it is easily seen that when $i = j$ both, $a$- and $b$-operators, obey Fermi-like relationships in two-dimensional sub-spaces, for example:

$$a_{+}a^{+}_{+} + a^{+}_{+}a_{+} = \begin{pmatrix} 1 & 0 & 0 & 0 \\ 0 & 1 & 0 & 0 \\ 0 & 0 & 0 & 0 \\ 0 & 0 & 0 & 0 \end{pmatrix} \quad a_{-}a^{+}_{-} + a^{+}_{-}a_{-} = \begin{pmatrix} 1 & 0 & 0 & 0 \\ 0 & 0 & 0 & 0 \\ 0 & 0 & 1 & 0 \\ 0 & 0 & 0 & 0 \end{pmatrix} \quad bb^{+} + b^{+}b = \begin{pmatrix} 1 & 0 & 0 & 0 \\ 0 & 0 & 0 & 0 \\ 0 & 0 & 0 & 0 \\ 0 & 0 & 0 & 1 \end{pmatrix}$$

However, $a^{+}_{-}b \neq \pm b\,a^{+}_{-}$ because $ba^{+}_{-} = 0$, but $a^{+}_{-}b \neq 0$, etc.

---